\documentclass[12pt]{article}
\usepackage{epsf,amssymb,amsmath,latexsym}
\title{Upper bounds on wavepacket spreading \\ for random Jacobi matrices}

\author{Svetlana Jitomirskaya$^1$, Hermann Schulz-Baldes$^{2}$
\\
\\
$^1$
{\small Department of Mathematics, University of California at Irvine, 
Ca, 92697, USA}
\\
$^2$
{\small Mathematisches Institut, Universit\"at
Erlangen-N\"urnberg, Germany}
}

\date{ }

\newtheorem{theo}{Theorem}
\newtheorem{defini}{Definition}
\newtheorem{proposi}{Proposition}
\newtheorem{lemma}{Lemma}
\newtheorem{coro}{Corollary}

\def\be{\begin{equation}}
\def\ee{\end{equation}}
\newcommand{\beq}{\begin{eqnarray}}
\newcommand{\eeq}{\end{eqnarray}}

\newcommand{\CC}{{\mathbb C}}
\newcommand{\NN}{{\mathbb N}}
\newcommand{\RR}{{\mathbb R}}

\newcommand{\ZZ}{{\mathbb Z}}

\newcommand{\PP}{{\bf P}}

\newcommand{\EE}{{\bf E}}

\newcommand{\Oo}{{\cal O}}

\newcommand{\Tt}{{\cal T}}

\newcommand{\Nn}{{\cal N}}

\newcommand{\comm}[1]{}
\newcommand{\comment}[1]{}

\begin{document}

\maketitle

\begin{abstract}
A method is presented for proving upper bounds on the
moments of the position operator when the dynamics of quantum
wavepackets is governed by a random (possibly correlated) Jacobi
matrix. As an application, one obtains sharp upper bounds on the
diffusion exponents for random polymer models, 
coinciding with the lower bounds obtained in a prior work. The
second application is an elementary argument 
(not using multiscale analysis or the Aizenman-Molchanov method) 
showing that under the condition of uniformly positive Lyapunov
exponents, the moments
of the position operator grow at most logarithmically in time.
\end{abstract}


\section{Introduction}

One of the fundamental questions of quantum mechanics concerns the spreading
of an initially localized wave packet $\phi$ under the time evolution 
$e^{-\imath tH}$ associated to a Schr\"odinger operator $H$. If the physical
space is $\RR^d$ or $\ZZ^d$ and the position operator is denoted by
$X$, the spreading can be quantified using the time-averaged moments of $X$
(or equivalently the moments of the 
associated classical probability distribution):
\begin{equation}
\label{eq-moment0}
M^q_T
\;=\;
\int^\infty_{0}\frac{dt}{T/2}\;e^{-\frac{2t}{T}}
\langle \phi |
\,e^{\imath H t}\,|X|^q\, e^{-\imath H t}\,
|\phi\rangle
\mbox{ , }
\qquad
q>0
\mbox{ . }
\end{equation}
It is well known that for short-range operators $H$, the moments cannot grow
faster than ballistically, that is $M^q_T\leq C(q)\,T^q$. 
The growth actually is
ballistic in typical scattering situations and 
for periodic operators $H$ describing Bloch electrons. 
On the other hand, if the moments $M^q_T$ are bounded uniformly in time, one
speaks of dynamical localization. This can be proven in the regime of Anderson
localization for random operators, but
also for certain almost-periodic operators 
(see \cite{Jit} for a review). 
There are many models where the moments $M^q_T$ exhibit some
non-trivial power law behavior. If $M^q_T\sim T^{q/2}$, the quantum motion is
diffusive, and any other asymptotic growth behavior is called anomalous
diffusion.  In order to distinguish various 
anomalous diffusive motions, one defines the diffusion exponents
\begin{equation} 
\label{ge}
\beta^+_q
\;=\;
\limsup_{T\to \infty}\;\frac{\log (M^q_T)}{\log(T^q)}
\;,
\qquad
\beta^-_q
\;=\;
\liminf_{T\to \infty}\;\frac{\log (M^q_T)}{\log(T^q)}
\;,
\qquad
q>0\;.
\end{equation}
If the limit exists, we write
$\beta_q =\beta_q^+=\beta_q^-$. The diffusion exponents 
correspond to the Levy-Khinchin classification of Levy flights in classical
probability, however, we stress that the quantum anomalous diffusion
does not result from a probabilistic dynamics, but rather from a
Hamiltonian one. It is due to delicate quantum interference
phenomena. The ballistic
bound implies $0\leq \beta_q^\pm\leq 1$ and convexity inequalities show that
$\beta_q^\pm$ is non-decreasing in $q$. In the regime of dynamical localization
$\beta_q=0$ and for quantum diffusion  $\beta_q=\frac{1}{2}$.
Anomalous diffusion corresponds to all other values of
$\beta_q^\pm$. Typically $\beta_q^\pm$ is then also varying with $q$
which reflects a rich
multiscale behavior of the wave packet spreading. Such anomalous
diffusion was exhibited numerically 
in several almost-periodic Jacobi matrices having singular continuous spectra
({\sl e.g.} Fibonacci and critical Harper operator), and also some
random and sparse Jacobi matrices. 

\vspace{.2cm}

It is a challenging problem of mathematical physics to calculate the diffusion
exponents for a given Schr\"odinger operator, in particular, when the quantum 
motion is anomalous diffusive. In this work we accomplish this for the 
so-called random polymer models, a random Jacobi matrix 
described in detail in the next section,  
and show that
\begin{equation}
\label{eq-diffequ}
\beta_q
\;=\;
\max\left\{\,0\,,\,1-\frac{1}{2q}\,\right\}
\;,
\qquad 
q>0\;,
\end{equation}
see Theorem \ref{theo-upper}.
This result confirms the heuristics and numerical results of Dunlap, Wu and
Phillips \cite{DWP} for the random dimer model, the prototype of a
the random polymer. The latter model was introduced and analyzed
in our prior work in collaboration with G.
Stolz \cite{JSS}, which already contained a rigorous proof of the
lower bound $\beta_q^-\geq 1-\frac{1}{2q}$. In this work we hence 
focus on the upper bound,
which amounts to proving quantitative localization estimates. 

\vspace{.2cm}

Next let us discuss this result in the context of prior rigorous 
work on other one-dimensional models (Jacobi matrices) 
exhibiting anomalous diffusion. 
First of all, the
Guarneri bound \cite{Gua} and its subsequent improvements
\cite{Com,Las,GSB2,GSB3,KL,BGT} allow to estimate  the diffusion
exponents from below in terms of various fractal dimensions of the spectral
measure. However, those results do not
allow to prove the lower bound in
\eqref{eq-diffequ} because, as was shown by de Bievre and Germinet \cite{BG}, 
the
random dimer model has pure-point spectrum so that the Hausdorff dimension
vanishes and 
the Guarneri bound is empty. In fact, the argument in \cite{JSS}
is based on a large deviation estimate on the localization length 
of the eigenstates near the so-called critical energies at which the Lyapunov
exponent vanishes. 

\vspace{.2cm}

Upper bounds on anomalous 
quantum diffusion were first proven for Jacobi matrices with
self-similar spectra \cite{GSB1,BS}, and these bounds are even optimal for the
so-called Julia matrices. 
Kiselev, Killip and Last \cite{KKL} presented a technique based on subordinacy
theory allowing to 
control the spread of a certain portion of the wave packet (not the fastest
one and therefore not the moments), and applied it to the Fibonacci model. 
Tcheremchantsev \cite{Tch} proved tight upper bounds on growing sparse 
potential Hamiltonians introduced in \cite{JL} and further analyzed in 
\cite{CM}. 
Recently, Damanik and Tcheremchantsev \cite{DT} developed a
transfer matrix based method that allows to prove upper bounds on the
diffusion exponents and also applied it to the Fibonacci
model. Another way to achieve upper bounds on the diffusion exponents
in terms of properties of the finite size approximants 
(Thouless widths and eigenvalue clustering)
was recently developed and applied to the
Fibonacci operator by Breuer, Last, and
Strauss \cite{BLS}. In the models considered in \cite{GSB1,BS,DT,BLS,Tch} 
the anomalous diffusion is closely linked to dimensional properties of the
spectral measures, even though the
(generalized) eigenfunctions have to be controlled as well. As a result, 
the transport slows
down as the fractal dimension of the spectral measure decreases. The origin
of the anomalous transport in the random polymer model  is of different
nature. In fact, in the random polymer model only a
few, but very extended localized 
states near the critical energies lead to the growth of
the moments. 
Hence this model illustrates that spectral theory may be of
little use for the calculation of the diffusion exponents. This statement is
even more true if the dimension of physical space is higher. There are
examples of three-dimensional operators with absolutely continuous spectral
measures, but subdiffusive quantum diffusion with diffusion exponents as low
as $\frac{1}{3}$ \cite{BeS}.

\vspace{.2cm}

The strategy for proving upper bounds advocated in \cite{DT}
appears to be more efficient in the present context 
than prior techniques \cite{GSB1,KKL}. We refine and generalize the
relevant part
in Section~\ref{sec-uppergen}, see in particular
Proposition~\ref{prop-DT}. It allows to give a rather simple proof 
of our second result, namely Theorem~\ref{theo-logbound}, which establishes 
a logarithmic bound $M^q_T\leq \log(T)^{\beta q}$, $\beta>2$, 
under the condition of uniformly
positive Lyapunov exponent. This result is neither new nor fully
optimal. However, in the generality we have (any condition on the distribution
of randomness, including {\it e.g.} Bernoulli) the Aizenman-Molchanov method
\cite{AM} cannot be applied, 
and the only technique previously available to obtain this statement was
the multi-scale analysis of \cite{CKM} (see also \cite{BG,DSS}). 
Thus our proof is significantly simpler. 
Moreover, one can argue that it captures the physically relevant effect of
localization. Indeed, it was shown by Gordon \cite{Gor} and del Rio, Makarov
and Simon \cite{DMS} that a generic rank one perturbation of a model in the
regime of strict dynamical localization ($M^q_T$ bounded) leads to
singular continuous spectrum, and therefore, by the RAGE theorem, growth of
the moments. However, it was shown in \cite{DJLS} that this growth can be at
most logarithmic, just as proven in Theorem~\ref{theo-logbound}. The
proof of Theorem~\ref{theo-logbound} constitutes essentially a part of the
proof of Theorem~\ref{theo-upper}.

\vspace{.2cm}

In the next section we present our models and results with technical details. 
Section~\ref{sec-uppergen} contains a general (non-random) strategy for
proving the upper bounds.  
Section \ref{sec-logbound} provides the proof of Theorem
\ref{theo-logbound} as well as some statements that are used in Section
\ref{5}. In Section \ref{4} we obtain probabilistic bounds on the transfer
matrices near a critical energy based on the large deviation estimate of
\cite{JSS}. Section \ref{5} contains the proof of Theorem \ref{theo-upper},
that is the identity \eqref{eq-diffequ}.

\vspace{.2cm}

\noindent {\bf Acknowledgment:} This work would have been impossible without
\cite{JSS}. We are very thankful to G. Stolz for this
collaboration. We also thank the anonymous referees for comments that
improved the paper.
The work of S. J. was supported in part by the NSF, grant DMS-0300974,
and  Grant No. 2002068 from the United States-Israel 
Binational Science Foundation (BSF), Jerusalem, Israel. 
H. 
S.-B. acknowledges support by the DFG.

\section{Models and results}

A Jacobi matrix is an operator $H_\omega$ on $\ell^2(\ZZ)$ associated to 
the data $\omega=(t_n,v_n)_{n\in\ZZ}$ of positive numbers $t_n$ and real
numbers $v_n$ which we suppose to be both bounded by a constant $C,$ and $t_n$
bounded away from $0$.   
Using the Dirac notation $|n\rangle$ for the canonical basis in
$\ell^2(\ZZ)$, $H_\omega$ is given by
\begin{equation} 
\label{oper}
H_\omega\,|n\rangle
\;=\;
t_{n+1}\,|n+1\rangle
\;+\;
v_{n}\,|n\rangle
\;+\;
t_{n}\,|n-1\rangle
\;.
\end{equation}
Each $\omega$ is  called a configuration. The set of all
configurations is contained in 
$\Omega=([-C,C]^{\times 2})^{\times\ZZ}$. 
The left shift $S$ is naturally  defined on $\Omega.$ 
A stochastic Jacobi matrix is a family $(H_\omega)_{\omega\in\Omega}$ of Jacobi
matrices drawn according to a probability measure $\PP$
on $\Omega$ which is invariant and ergodic w.r.t. to $S$. Furthermore,
we speak of a random Jacobi matrix if $\PP$ has at most finite distance
correlations, namely there exists a finite correlation length $L\in\NN$ 
such that $(t_n,v_n)$ and 
$(t_m,v_m)$ are independent 
whenever $|n-m|\geq L$. Most prominent example of a random Jacobi
matrix is the one-dimensional 
Anderson model for which $t_n=1$ and the $v_n$ are independent and identically
distributed so that $L=1$. Random polymer models as studied in \cite{DWP,JSS}
and described in more detail below
provide an example of a random Jacobi matrix with
finite distance correlations (the second crucial feature of these models is
that the $(t_n,v_n)$ only take a finite number of values). 

\vspace{.2cm}

We will consider the disorder and time averaged moments  
of the position operator $X$ on $\ell^2(\ZZ)$, denoted by
$M^q_T$  as in \eqref{eq-moment0}:
\begin{equation}
\label{eq-moment}
M^q_T
\;=\;
\int^\infty_{0}\frac{dt}{T/2}\;e^{-\frac{2t}{T}}
\;\EE\;\langle 0 |
\,e^{\imath H_\omega t}\,|X|^q\, e^{-\imath H_\omega t}\,
|0\rangle
\mbox{ , }
\qquad
q>0
\mbox{ . }
\end{equation}
Here $\EE$ denotes 
the average over $\omega$ w.r.t. $\PP$. (Note that upper bounds on the
expectation w.r.t. $\PP$ yield upper bounds almost surely.) 
One may replace $|0\rangle$ by any other localized initial state (at least in
$\ell^1(\ZZ)$). 

\vspace{.2cm}

As discussed in the introduction, 
it is well-known that the Anderson model exhibits 
dynamical localization, that is $M^q_T\leq C(q)<\infty$ uniformly in $T$ for
all $q>0$. 
A by-product of our analysis of the random polymer model discussed below is
a simple proof of the weaker result that $M^q_T$ grows 
at most logarithmically in $T$ whenever the Lyapunov exponent is strictly
positive. In order to define the latter, let us introduce
as usual in the analysis of one-dimensional systems 
the transfer matrices at a complex energy $z$ by
$$
\Tt^z_\omega(n,m)
\;=\;
\Tt^z_{n-1}\cdot\ldots\cdot\Tt^z_{m}\;,
\qquad n> m\;,
\qquad
\Tt^z_n
\;=\;
\left(\begin{array}{cc} (z-v_n)t_n^{-1} & -t_n \\ t_n^{-1} & 0 
\end{array} \right)
\;.
$$
Furthermore $\Tt^z_\omega(n,m)=\Tt^z_\omega(m,n)^{-1}$ for $n<m$ and 
$\Tt^z_\omega(n,n)={\bf 1}$.
Then the Lyapunov exponent is 
$$
\gamma(z)
\;=\;
\lim_{N\to\infty}\;\frac{1}{N}\;
\EE\;\log\left(\,\|\Tt^z_\omega(N,0)\|\,\right)
\;.
$$
Because the $t_n$ and $v_n$ are uniformly bounded, one shows by  
estimating the norm of a
product of matrices by the product of their norms  that
for every bounded set $U\subset \CC$
\begin{equation}
\label{eq-uniformupper}
\|\Tt_\omega^z(n,m)\|
\;\leq\;
e^{\gamma_1\,|n-m|}
\mbox{ , }
\qquad z\in U\;,
\end{equation}
where the $\gamma_1$ depends on $U$. This implies that
$\gamma(z)\leq \gamma_1$ for $z\in U$. For a random Jacobi matrix, 
uniform lower bounds on $\gamma(z)$
can be proven by the Furstenberg theorem
({\it e.g.} \cite{PF}). 
This applies in particular to the Anderson model (also with
a Bernoulli potential), and more generally to random Jacobi matrices with
 correlation length equal to $1.$

\begin{theo}
\label{theo-logbound}
Consider a random Jacobi matrix.
Let the {\rm (}non-random{\rm )}  
spectrum be $\sigma(H)\subset (E_0,E_1)$ 
Suppose that
the Lyapunov exponent is strictly positive:
\begin{equation}
\label{eq-lyaplower}
\gamma(z)
\;\geq\;
\gamma_0\;>\;0\;,
\qquad
z\in (E_0,E_1)\;.
\end{equation}
Then for any $\beta>2$ there exists a constant $C(\beta,q)$ such that
\begin{equation}
\label{eq-logupper}
M_T^q
\;\leq\;
(\log T)^{q\beta}\;+\;C(\beta,q)
\mbox{ . }
\end{equation}

\end{theo}

Another self-averaging quantity associated to to ergodic Jacobi
matrices is the integrated density of states (IDS) $\Nn(E)$ which can
be defined by
$$
\int \Nn(dE)
\,f(E)
\;=\;
\EE\; 
\langle 0|f(H_\omega)|0\rangle
\;,\;
\qquad
f\in C_0(\RR)
\;.
$$
The Thouless formula ({\it e.g.} \cite{PF}) links the Lyapunov exponent
to the IDS. It implies that, if \eqref{eq-lyaplower} holds for real
$E\in(E_0,E_1)$, then one also has $\gamma(z)\geq\gamma_0$ for all $z$
with $\Re e (z)\in(E_0,E_1)$.   

\vspace{.2cm}

As shown by Theorem~\ref{theo-logbound} and the remark just before it,
it is necessary for a random Jacobi matrix to have a 
correlation length larger than $1$ for the moments $M_T^q$ to grow faster than
logarithmically, so that
the diffusion exponents $\beta_q$ do not vanish.
That this actually happens for
the random dimer model was discovered by Dunlap, Wu and Phillips \cite{DWP}.

\vspace{.2cm}

Next let us describe in more detail 
the more general random polymer model considered in \cite{JSS}.
Given are two finite sequences
$\hat{t}_\pm=(\hat{t}_\pm(0),\ldots,\hat{t}_\pm(L_\pm-1))$ and
$\hat{v}_\pm=(\hat{v}_\pm(0),\ldots,\hat{v}_\pm(L_\pm-1))$ 
of real numbers, satisfying
$\hat{t}_\pm(l)>0$ for all $l=0,\ldots, L_\pm-1,\;  L_\pm\geq 1$. 
The associated random polymer model is the random Jacobi matrix
constructed by random juxtaposition
of these sequences and randomizing the origin. More precisely,
configurations $\omega\in\Omega$ can be identified with the data  
of a sequence of signs
$(\sigma_n)_{n\in\ZZ}$ and an integer $0\leq l \geq L_{\sigma_1}-1$, via the
correspondence 
$(t_n)=
(\ldots,\hat{t}_{\sigma_1}(l),\ldots, \hat{t}_{\sigma_1}(L_{\sigma_1}-1), 
\hat{t}_{\sigma_2}(0),\ldots,
\hat{t}_{\sigma_2}(L_{\sigma_2}-1),\hat{t}_{\sigma_3}(0),\ldots)$ 
and similarly for $(v_n),$ with choice of origin
$t_0=\hat{t}_{\sigma_1}(l)$ and $v_0=\hat{v}_{\sigma_1}(l)$. The shift is as usual, and the
probability $\PP$ is 
the Bernoulli measure with
probabilities $p_+$ and $p_-=1-p_+$ combined with a randomization for $l$
({\it cf.} \cite{JSS} for details). The correlation length in this model is
$L=\max\{L_+,L_- \}$.
It is now natural and convenient to consider the
polymer transfer matrices 
\begin{equation}
\label{eq-transfer}
T^z_\pm
\;=\;
\Tt^z_{\hat{v}_\pm(L_{\pm}-1),\hat{t}_\pm(L_{\pm}-1)} 
\cdot\ldots\cdot 
\Tt^z_{\hat{v}_\pm(0),\hat{t}_\pm(0)}
\mbox{ , }
\qquad \mbox{where }\;\;\;\;\;
\Tt^z_{v,t}\;=\;
\left(\begin{array}{cc} (E-v)t^{-1} & -t \\ t^{-1} & 0 \end{array} \right)
\mbox{ . }
\end{equation}

\begin{defini}
\label{def-critical}
An energy $E_c\in\RR$ is called critical for the random polymer model
$(H_\omega)_{\omega\in\Omega}$ if
the polymer transfer matrices  $T_\pm^{E_c}$ are elliptic {\rm (}i.e.
$|\mbox{\rm Tr}(T_\pm^{E_c})|<2${\rm )} or equal to $\pm{\bf 1}$ and
commute
\begin{equation}
\label{eq-critical}
[T^{E_c}_-,T^{E_c}_+]\;=\;0\mbox{ . }
\end{equation}
\end{defini}

If $L_\pm=1$, the model reduces to the
Bernoulli-Anderson model and there are no critical energies. 
The most studied \cite{DWP,Bov,BG} example is
the random dimer model for which $L_+=L_-=2$ and 
$\hat{t}_\pm(0)=\hat{t}_\pm(1)=1$,
$\hat{v}_+(0)=
\hat{v}_+(1)=\lambda$ and
$\hat{v}_-(0)=\hat{v}_-(1)=-\lambda$  for some $\lambda\in\RR$.
This model has two critical energies $E_c=\lambda$
and $E_c=-\lambda$ as long as $\lambda<1$. For further examples we refer to
\cite{JSS}. It follows from the definition that a simultaneous change
of coordinates reduces both $T_+$ and $T_-$ to rotations by angles
that are denoted by $\eta_+$ and $\eta_-$.

\vspace{.2cm}

It is immediate from \eqref{eq-critical} that the Lyapunov exponent vanishes
at a critical energy. Because the transfer matrices $T_\pm^z$ are analytic in
$z$, it follows that there is a constant $c_0$ such that for all
$\epsilon\in\CC$ with $|\epsilon|<\epsilon_0$ one has for $n,m\in \NN$,
\begin{equation}
\label{eq-lyapbound}
\left\|
{\cal T}^{E_c+\epsilon}_{\omega}(n,m)\right\|
\;\le \;
e^{c_0\,|\epsilon|\,|n-m|}
\mbox{ . }
\end{equation}
\noindent In particular, 
$|\gamma(E_c+\epsilon)|\leq \,c_0\,|\epsilon|$. However, the correct
asymptotics for the Lyapunov exponent is
$\gamma(E_c+\epsilon)=\Oo(\epsilon^2)$. This was first shown
(non-rigorously) by Bovier for the case of the random dimer model \cite{Bov},
but heuristics were already given in \cite{DWP}.
The rigorous result about the Lyapunov exponent and also the integrated
density of states  $\Nn$ 
are combined in the following theorem. 

\begin{theo}
\label{theo-lyap} {\rm \cite{JSS}}
Suppose that $\EE( e^{2\imath\eta_\sigma}) \neq 1$ and
$\EE( e^{4\imath\eta_\sigma})\neq 1$. Then for $\epsilon\in\RR$ and some
$D\geq 0$, the Lyapunov exponent of a random polymer model satisfies
\begin{equation}
\label{eq-lyap}
\gamma(E_c+\epsilon)
\;=\;
D\,\epsilon^2\;+\;
\Oo(\epsilon^3)
\;,
\end{equation}
in the vicinity of a critical energy $E_c$.
If $\|\,[T^{E_c+\epsilon}_-,T^{E_c+\epsilon}_+]\,\|
\geq C\,\epsilon$ for some $C>0$ and small $\epsilon$, one has $D>0$. 
Moreover, the {\rm IDS} $\Nn$ satisfies 
\begin{equation}
\label{eq-DOS}
\Nn(E_c+\epsilon)-\Nn(E_c-\epsilon)
\;=\;
D'\,\epsilon
\;+\;
\Oo(\epsilon^2)
\;,
\end{equation}
for some constant $D'>0$.
\end{theo}

Furthermore \cite{JSS}
contains explicit formulas for $D$ and $D'$.
The bound $D>0$ is not explicitely contained in \cite{JSS}, but can be
efficiently checked using Proposition~1 in \cite{SSS}. The statement
about the IDS only requires $\EE( e^{2\imath\eta_\sigma}) \neq 1.$ 
Let us remark that the hypothesis
$\EE( e^{4\imath\eta_\sigma}) \neq 1$ does not hold, for example, 
in the special case of a random dimer model if $\lambda=1/\sqrt{2}$. In this
situation, one is confronted with an anomaly. Nevertheless, the
asymptotics is as in \eqref{eq-lyap} and again one can calculate $D$
explicitely \cite{Sch}. As we did not perform the large deviation analysis of
\cite{JSS} in the case of an anomaly, we retain the hypothesis of
Theorem~\ref{theo-lyap} throughout. 

\vspace{.2cm}

The following is the main result
of this work. 

\begin{theo}
\label{theo-upper}
Suppose that $\EE( e^{2\imath\eta_\sigma}) \neq 1$ and
$\EE( e^{4\imath\eta_\sigma})\neq 1$, and that the random polymer model has a
critical energy at which {\rm \eqref{eq-lyap}} holds with $D>0$. 
Then for $q>0$,
$$
\beta_q\;=\;\max\left\{\,0\,,\,1-\frac{1}{2q}\,\right\}
\;.
$$
\end{theo}

As already pointed out, the lower bound $\beta_q^-\geq 1-\frac{1}{2q}$ was
already proven in \cite{JSS}.

\section{A strategy for proving upper bounds on dynamics}
\label{sec-uppergen}

In this section it is not necessary for the Jacobi matrix to be random
or ergodic; hence
the index $\omega$ and the average $\EE$ are suppressed. Let the notation for
the Green's function be
\begin{equation}
\label{eq-green}
G^{z}(n,m)
\;=\;
\langle n|\,\frac{1}{H-z}\,|m\rangle
\mbox{ , }
\end{equation}
where $n,m\in\ZZ$ and $z\in\CC$ is not in the spectrum $\sigma(H)$ of $H$.
The starting point of the analysis is to express
the time averaged moments \eqref{eq-moment} 
in terms of the Green's function
\begin{equation}
\label{eq-greentransport}
M^q_T
\;=\;
\sum_{n\in \ZZ}
|n|^q\;\int_\RR \frac{dE}{\pi\,T}
\;|G^{E+\frac{\imath}{T}}(0,n)|^2
\mbox{ . }
\end{equation}
Using the spectral theorem, this well known identity can be checked 
immediately by a
contour integration. In order to decompose the expression on the r.h.s., let
us introduce for $0<\alpha_0<\alpha_1$ and $E_0\leq E_1$,
\begin{equation}
\label{eq-greenpart}
M^{q,\alpha_0,\alpha_1}_T(E_0,E_1)
\;=\;
\sum_{T^{\alpha_0}
< |n|\leq T^{\alpha_1}}
|n|^q\;\int_{E_0}^{E_1} \frac{dE}{\pi\,T}
\;|G^{E+\frac{\imath}{T}}(0,n)|^2
\mbox{ , }
\end{equation}
and $M^{q,0,\alpha}_T(E_0,E_1)$ is defined similarly with the sum
running over $0\leq |n|\leq T^\alpha$, $\alpha>0$.
The following result also holds for higher dimensional models.

\begin{proposi}
\label{prop-CT}
Suppose $\sigma(H)\subset (E_0,E_1)$ and $\alpha>1$. Set 
$\Delta=\mbox{\rm dist}(\{E_0,E_1\},\sigma(H))$.
Then there exists a constant
$C_1=C_1(\alpha,\Delta,q)$ such that
$$
\left|\,M^q_T
\,-\,M^{q,0,\alpha}_T(E_0,E_1)\,\right|
\;\leq\;
\frac{C_1}{T}
\;.
$$
\end{proposi}

The proof is  based on the following Combes-Thomas estimate. 
Even though standard, its proof is sufficiently short and beautiful to
reproduce it.

\begin{lemma}
\label{lem-CT}
Let $\Delta(z)=\mbox{\rm dist}(z,\sigma(H))$ and $C_2=(4\|t\|_\infty)^{-1}$  
where $\|t\|_\infty=\sup_{n\in\ZZ}t_n$. Then
$$
|\,G^z(n,m)\,|
\;\leq\;
\frac{2}{\Delta(z)}\;
\exp\left(\,-\,\mbox{\rm arcsinh}(C_2\,\Delta(z))\,
|n-m|
\right)
\;.
$$
\end{lemma}

\noindent {\bf Proof.} For $\eta\in\RR$, 
set $H_\eta=e^{\eta X}He^{-\eta X}$. A
short calculation shows that
$$
\|\,H_\eta-H\,\|
\;\leq\;
\|t\|_\infty\,|e^\eta-e^{-\eta}|
\;.
$$
%
Hence one has
$$
\left\|(H_\eta-z)^{-1}\right\|
\;\leq\; 
\left(\,\left\|(H-z)^{-1}\right\|^{-1}\,-\,\|H_\eta-H\|\,\right)^{-1}.
$$
Since $\|(H-z)^{-1}\|\leq \Delta(z)^{-1}$, the choice $|\eta|=
\mbox{\rm arcsinh}\left(\Delta(z)/(4\,\|t\|_\infty)\right)$
hence gives $\|(H_\eta-z)^{-1}\|\leq 2/\Delta(z)$.
The bound now follows from $\langle n|(H-z)^{-1}|m\rangle
=e^{\eta(m-n)}\,\langle n|(H_\eta-z)^{-1}|m\rangle$.
\hfill $\Box$

\vspace{.2cm}

The following estimate will be used not only for
the proof of Proposition~\ref{prop-CT}, but 
at several reprises below.

\begin{lemma}
\label{lem-sumbound}
Let $\Delta,\alpha>0$, $q\geq 0$ and $N\in\NN$. Let
$p=[\frac{q+1}{\alpha}]$ where $[\,\cdot\, ]$ stands for the integer part.
Then
$$
\sum_{n > N}\,n^q\,e^{-\Delta\,n^\alpha}
\;\leq\;
 \frac{2}{\alpha}\;p!\;\bigl(N+\Delta^{-1}\bigr)^{p}\;
\frac{e^{-\Delta N^\alpha}}{\Delta}
\;.
$$
\end{lemma}

\noindent {\bf Proof.}
Bounding the sum by the integral on each interval of monotonicity and
then extending this integral to the entire range $[N,\infty)$, we obtain
$$
\sum_{n > N}\,n^q\,e^{-\Delta\,n^\alpha}
\;\leq\;
2\int_N^\infty dx\;x^{q}\,e^{-\Delta x^\alpha}
\;\leq\;
\frac 1\alpha\,\int_{N^{\alpha}}^\infty dy
\;y^p\,e^{-\Delta y}
\;=\;
\frac{\,e^{-\Delta N^\alpha }}{\alpha\,\Delta^{1+p}}
\;
\sum_{j=0}^{p}\frac{p!}{j!}\,(N\Delta)^j
\;.
$$
Bounding the sum over $j$ by $p!\,(N\Delta+1)^{p}$ completes the proof.
\hfill $\Box$

\vspace{.2cm}

\noindent {\bf Proof} of Proposition~\ref{prop-CT}.
We first consider the energies above the spectrum and set $\Delta=
\mbox{\rm dist}(E_1,\sigma(H))$.
Due to the previous two lemmata and if $p=[q+1]$,
\begin{eqnarray*}
M^{q,0,\infty}_T(E_1,\infty)
& \leq &
\int^\infty_0\frac{dE}{\pi\, T}\;
\sum_{|n|\geq 1}\,|n|^q\,\frac{2}{\Delta+E}
\;
\exp\left(-\mbox{\rm arcsinh}(C_2(\Delta+E))|n|\right)\,
\\
& \leq &
\int^\infty_0\frac{dE}{\pi \,T}
\;
\frac{8\,p!}{\Delta+E}
\;\;
\mbox{\rm arcsinh}(C_2(\Delta+E))^{-(p+1)}
\;
\;.
\end{eqnarray*}
Since $\mbox{\rm arcsinh} (y)\geq \ln (y)$ for large $y,$ this shows that
$M^{q,0,\infty}_T(E_1,\infty)\leq C /T$ for some constant
$C=C(q,\Delta)$. A similar bound holds for $M^{q,0,\infty}_T(-\infty,E_0)$.
Now using the imaginary part of the energy in Lemma~\ref{lem-CT} and the bound
$\mbox{\rm arcsinh}(y)\geq y$ for sufficiently small $y\geq 0$, 
we obtain
\begin{eqnarray*}
M^{q,\alpha,\infty}_T(E_0,E_1)
& \leq &
\frac{2}{\pi}\;
\int^{E_1}_{E_0}dE\;
\sum_{|n| > T^{\alpha}}\,|n|^q
\;
\exp\left(-\mbox{\rm arcsinh}(C_2/T)|n|\right)\,
\\
& \leq &
\frac{4\,(E_1-E_0)\,p!}{\pi}\;\;
\frac{e^{-C_2T^{\alpha-1}}}{C_2/T}
\;
\left(T^{\alpha}+T/C_2\right)^{p}
\;.
\end{eqnarray*}
For $\alpha>1$, this decreases faster than any power of $T$.
Combining these estimates implies the proposition.
\hfill $\Box$

\vspace{.2cm}

According to Proposition~\ref{prop-CT} and equation 
\eqref{eq-greenpart}, one now needs a good bound on the decay (in $n$) of the 
Green's function for
complex energies in the vicinity of the spectrum. 
As shown by  Damanik and Tcheremchantsev \cite{DT}, such bounds can be
obtained for Jacobi matrices in a very efficient way in 
terms of the transfer matrices. 
Here we give a streamlined  proof of this statement which also works for
arbitrary Jacobi matrices (the kinetic part is not necessarily the discrete
Lapalacian) and does not contain energy dependent constants as in \cite{DT}. 
It will be convenient to first consider such bounds for the half-line
problem, which is operator \eqref{oper} on $\ell^2(\NN)$ with
Dirichlet boundary conditions.
This operator and its Green's function are denoted by
$\hat H$ and $\hat G^z(n,m)$.

\begin{proposi}
\label{prop-DT}
Set $\tau=\max\{\|t\|_\infty^2\,,\,1\,,\,\|t^{-1}\|_\infty^2\}$ and
$z=E+\frac{\imath}{T}$. One has the bounds
$$
\sum_{n> N}
\;|\hat G^z(0,n)|^2
\;\leq\;
\frac{4\,\tau^3\;T^4}{
\;\max_{0\leq n \leq N}\;
\|\,\Tt^z(n,0)\,\|^2\;}
\;,
$$
and, for $T\geq 1$,
$$
\sum_{|n|> N}
\;|G^z(0,n)|^2
\;\leq\;
\frac{16\,\tau^4\;T^6}{
\;\max_{0\leq |n| \leq N}\;
\|\,\Tt^z(n,0)\,\|^2\;}
\;.
$$
\end{proposi}

\noindent {\bf Proof.} 
Let $\Gamma_N$ be the decoupling operator at
$N$ defined by $\Gamma_N=t_{N+1}(|N\rangle\langle N+1|+|N+1\rangle\langle N|)$
and set $\hat G^z_N  = (\hat H-z-\Gamma_N)^{-1}$ and 
$\hat G^z= (\hat H-z)^{-1}$.
The resolvent identity reads
$$ 
\hat G^z
\;=\;
\hat G^z_N\,-\,\hat G^z_N\,\Gamma_N \,\hat G^z
\;.
$$
Thus, with the notation $\hat G^z_N(n,m)=\langle n|\hat G^z_N|m\rangle$,
\begin{eqnarray*}
\displaystyle\sum_{n>N}|\hat G^z(0,n)|^2
& = & 
\displaystyle\sum_{n>N}|\hat G^z_N\Gamma_N \hat G^z(0,n)|^2
\\
& = &
|\hat G^z_N(0,N)t_{N+1}|^2
\displaystyle\sum_{n>N}|\hat G^z(N+1,n)|^2
\;\leq\; 
\|t\|_\infty^2 T^{2}
|\hat G^z_N(0,N)|^2
\;,
\end{eqnarray*}
since $T^{-1} = \Im m(z)$ and $\|\hat G^z\|\leq T.$ As the l.h.s. is
decreasing in $N,$ we therefore have
\begin{equation} 
\label{eq-min}
\displaystyle\sum_{n>N}|\hat G^z(0,n)|^2
\;\leq \;
\|t\|_\infty^2\, T^{2}
\min_{0\leq n\leq N}|\hat G^z_n(0,n)|^2
\;.
\end{equation}

Now let 
$\Pi_N=\sum^{N}_{n=0}|n\rangle\langle n|$ be the projection on the states on
the first $N+1$ sites and set $\hat H_N=\Pi_NH\Pi_N$. 
Then $\hat G^z_N(n,m)$, $0\leq n,m\leq N$, 
are the matrix elements of the inverse of an $(N+1)\times (N+1)$ 
matrix $\hat H_N-z$. The matrix elements are  
closely linked to the transfer matrix 
$$
\Tt^z(N+1,0)
\;=\;
\left(\begin{array}{cc} a & b \\ c & d 
\end{array} \right)
\;,
$$
since $a,b,c,d$ 
when multiplied by the $\prod_{n=1}^N t_n,$ 
are the determinants of certain minors of $z-\hat H_N$. 
Namely by Cramer's rule (or, alternatively, by the Stieltjes continued
fraction expansion and geometric resolvent identity) 
the following identities hold for $z\notin \sigma(\hat H_N)$:
$$
\hat G^z_N(0,0)
\;=\;\frac{1}{t_0^2}\;\frac{b}{a}\;,
\qquad
\hat G^z_N(N,N)
\;=\;-\,\frac{c}{a}
\;,
\qquad
\hat G^z_{N-1}(0,0)\;=\;\frac{1}{t_0^2}\;\frac{d}{c}\;,
$$ 
and
$$
\hat G^z_N(0,N)
\;=\;-\,\frac{1}{t_0}\;\frac{1}{a}\;,
\qquad
\hat G^z_{N-1}(0,N-1)\;=\;-\,\frac{1}{t_0\,t_N}\;\frac{1}{c}\;.
$$
Therefore $|b|\leq t_0^2\,T\,|a|$, $|c|\leq T\,|a|$ and
$|d|\leq  t_0^2\,T\,|c|$. 
As the matrix norm is bounded by the Hilbert-Schmidt norm,
it follows that 
$$
\|\Tt^z(N+1,0)\|^2
\;\leq \;
\frac {4\,T^2\;\tau^2}
{\min\{|\hat G_N^z(0,N) |^2,|\hat G^z_{N-1}(0,N-1)|^2\}}\;.
$$ 
By (\ref{eq-min}) this proves the first inequality.
The second one follows from the first one (coupled with the same
statement for the left half-line) by observing that the
resolvent identity gives
$$
G^z(0,n)
\;=\;
\hat G^z(0,n)\,-\,G^z(0,-1)\,t_0\,
\hat G^z(0,n)
\;.
$$
Therefore $|G^z(0,n)|\leq (1+Tt_0)\, |\hat G^z(0,n)|$ which implies the second
bound. 
\hfill $\Box$

\vspace{.2cm}

\section{Logarithmic bounds in the localization phase}
\label{sec-logbound}

In this section we provide the proof of Theorem~\ref{theo-logbound} and hence
suppose throughout that the stated hypothesis hold. The main
idea is to use the given positivity of the Lyapunov exponent 
\eqref{eq-lyaplower}, combine it with the given uniform upper bound
\eqref{eq-uniformupper} in order to deduce good probabilistic estimates on the
growth of the transfer matrices. This growth in turn allows to bound the
Green's function due to Proposition~\ref{prop-DT} which then readily
leads to the logarithmic upper bound on the moments.

\vspace{.2cm}

Let us set  $U=\{z\in\CC\,|\,E_0\leq \Re e (z)\leq E_1\,,\, |\Im m
(z)|\leq 1\,\}$. 

\begin{lemma}
\label{lem-probboundsimple}
For $z\in U$ and $N\in\NN$, the set
$$
\Omega_N(z)
\;=\;
\left\{
\,\omega\in\Omega\;\left|\;
\|\Tt_\omega^z(N,0)\|^2\,\geq \,e^{\gamma_0\,N}
\;\right.
\right\}
$$
satisfies
$$
\PP(\Omega_N(z))
\;\geq\;
\frac{\gamma_0}{2\,\gamma_1-\gamma_0}
\;.
$$
\end{lemma}

\noindent {\bf Proof.}
Let us set $P=\PP(\Omega_N(z))$. Due to \eqref{eq-lyaplower}, the
subadditivity of the transfer-matrix cocycle
and the bound \eqref{eq-uniformupper}, it follows that
$$
\gamma_0\;\leq\;\frac{1}{N}\;\EE\;\log(\|\Tt_\omega^z(N,0)\|)
\;\leq\;
(1-P)\,\frac{1}{2}\,\gamma_0\;+\;P\,\gamma_1
\;,
$$
with $\gamma_1$ defined by \eqref{eq-uniformupper} using $U$ as above. This
directly implies the result.
\hfill $\Box$

\begin{lemma}
\label{lem-probboundsimple2}
Let $z\in U$ and $N\in\NN$. Then there is a constant 
$C_3=C_3(\gamma_0,\gamma_1)$ such that 
the set
$$
\hat{\Omega}_N(z)
\;=\;
\left\{
\,\omega\in\Omega\;\left|\;
\max_{0\leq n \leq N}\,\|\Tt_\omega^z(n,0)\|^2
\,\geq\, e^{C_3\,N^{\frac{1}{2}}}
\;\right.
\right\}
$$
satisfies
$$
\PP(\hat{\Omega}_N(z))
\;\geq\;
1\,-\,e^{-\,C_3\,N^{\frac{1}{2}}}
\;.
$$
\end{lemma}

\noindent {\bf Proof.}
Let us split $N$ into $\frac{N}{N_0}$ pieces of length $N_0$ (here and in the
sections below, we suppose without giving further details 
that there is an integer number of pieces and that the boundary terms
are treated separately). By the stationarity, on each piece 
$[jN_0+1,(j+1)N_0)$,
Lemma~\ref{lem-probboundsimple} with $N=N_0$ applies. As the pieces are 
independent, we deduce
$$
\PP\left(
\left\{
\,\omega\in\Omega\;\left|\;
\max_{0\leq j \leq N/N_0}\,\|\Tt_\omega^z((j+1)N_0,jN_0+1)\|^2
\,\leq\, e^{\gamma_0\,N_0}
\;\right.
\right\}
\right)
\;\leq\;
(1-p_0)^{\frac{N}{N_0}}
\;,
$$
where $p_0=\gamma_0/(2\,\gamma_1-\gamma_0)$. Furthermore
$\Tt_\omega^z((j+1)N_0,jN_0+1)=
\Tt_\omega^z((j+1)N_0,0)\Tt_\omega^z(jN_0,0)^{-1}$. 
As $A=BC$ implies either
$\|B\|\geq \|A\|^{\frac{1}{2}}$ or $\|C\|\geq \|A\|^{\frac{1}{2}}$ for
arbitrary matrices, and $\|A^{-1}\|= \,\|A\|$ for
$A\in\,$SL$(2,\CC)$, it therefore follows that
$$
\PP\left(
\left\{
\,\omega\in\Omega\;\left|\;
\max_{0\leq j \leq N/N_0}\,\|\Tt_\omega^z(jN_0,0)\|^2
\,\geq\, \;e^{\frac{1}{2}\,\gamma_0\,N_0}
\;\right.
\right\}
\right)
\;\geq\;
1\,-\;(1-p_0)^{\frac{N}{N_0}}
\;.
$$
Choosing $N_0=cN^{\frac{1}{2}}$ with adequate $c$ concludes the proof.
\hfill $\Box$

\vspace{.2cm}

Since the above lemma applies equally well to negative integers $N$,
it is a direct corollary of Lemma~\ref{lem-probboundsimple2} and
Proposition~\ref{prop-DT} that, for 
$T$ sufficiently large,
\begin{equation}
\label{eq-avgreenbound}
\sum_{|n|> N}
\;\EE\;|G_\omega^z(0,n)|^2
\;\leq\;
32\,\tau^4
\;T^6\,e^{-\,C_3\,N^{\frac{1}{2}}}\;,
\qquad
z\;=\;E\,+\,\frac{\imath}{T}\,\in\,U\;.
\end{equation}

\vspace{.1cm}

The following lemma, holding for arbitrary ergodic families of Jacobi
matrices, is useful for bounding
a sum similar to the one in \eqref{eq-avgreenbound}, but over
$|n|\leq N$. 
Let $B_\mu(z)=\int\mu(dE) \,(z-E)^{-1}$ denote the Borel transform of a
measure $\mu$. 

\begin{lemma}
For any $E\in\RR$, $T>0$ and $N\geq 1$, one has
\begin{equation} 
\label{m1a}
\sum_{0\leq |n|\leq N}\frac{\,|n|^q\;}{\pi\,T}
\;\EE\;|G_\omega^{z}(0,n)|^2
\;\leq \;
\frac {N^q}{\pi}\;\Im m\; 
B_\Nn(z)
\;,
\qquad
z\;=\;E\,+\,\frac{\imath}{T}
\;.
\end{equation}
Furthermore, for any $E_0<E_1$, 
\begin{equation}
\label{m1b} 
\sum_{0\leq |n|\leq N}\,|n|^q\;\int_{E_0}^{E_1}\frac{dE}{\pi\,T}
\;\EE\;|G_\omega^{E+\frac{\imath}{T}}(0,n)|^2
\;\leq\;
N^q\;.
\end{equation}
\end{lemma}

\noindent {\bf Proof.} One has
\begin{eqnarray*}
\sum_{0\leq |n|< N}\frac{\,|n|^q\;}{\pi\,T}
\;\EE\;|G_\omega^{z}(0,n)|^2
& \leq & 
N^q \sum_{n\in\ZZ}
\frac{1}{\pi\,T}\;\EE\;
\langle
0|\frac{1}{H_\omega-E-\frac{\imath}{T}}|n\rangle\; 
\langle n|\frac{1}{H_\omega-E+\frac{\imath}{T}}|0\rangle\;
\\ 
& \leq &
N^q\;
\EE\;
\frac{1}{\pi\,T}\; 
\langle 0|\frac{1}{(H_\omega-E)^2+\frac{1}{T^2}}|0\rangle
\;,
\end{eqnarray*}
which by the spectral theorem gives \eqref{m1a}. Using
$$
\int_{E_0}^{E_1}\frac{dE}{\pi}\;\Im m \;B_\Nn
(E+\imath\, T^{-1})
\;\leq\; 
\int_\RR \frac{dE}{\pi\, T}\,\int_\RR\Nn(de)\;
\frac{1}{(e-E)^2+\frac 1{T^2}}
\;=\;\int_\RR\Nn(de)
\;=\;1\;,
$$
the inequality 
\eqref{m1b} follows upon integrating \eqref{m1a}. 
\hfill $\Box$

\vspace{.2cm}

\noindent {\bf Proof} of Theorem~\ref{theo-logbound}.
According to Proposition~\ref{prop-CT} it remains to bound
$\EE\, M^{q,0,\alpha}_T(E_0,E_1)$ 
for $\alpha>1$. We further split this into two
contributions:  
$$
M_T(1)
\;=\;
\sum_{0\leq |n|\leq (\log T)^\beta}
\,|n|^q\;\int_{E_0}^{E_1} \frac{dE}{\pi\,T}
\;\EE\;|G_\omega^{E+\frac{\imath}{T}}(0,n)|^2
\;,
$$
and $M_T(2)$ corresponding to the sum over $(\log T)^\beta < |n|\leq T^{\alpha}$.
The bound
(\ref{m1b}) with $N=(\log T)^\beta$ immediately gives $M_T(1)\leq
(\log T)^{\beta q}$. The second contribution can be
bounded using \eqref{eq-avgreenbound}:
$$
M_T(2)
\;\leq\;
\frac{64\;\tau^4\;T^5}{\pi}
\sum_{n\geq (\log T)^\beta}\,n^q
\,e^{-\,C_3\,n^{\frac{1}{2}}}
\;,
$$
which is bounded by a constant $C(\beta,q)$ 
due to Lemma~\ref{lem-sumbound} as long as
$\beta>2$. 
\hfill $\Box$

\vspace{.2cm}

This above proof applies equally well to the half-line problem
with arbitrary boundary condition, the
only difference being that the IDS $\Nn$ has to be replaced by the
$\EE(\mu_\omega(dE))$ where $\mu_\omega$ is the spectral measure of
$|0\rangle$ and $\hat H$.
 
\section{Probabilistic estimates near a critical energy}
\label{4}

In this section, we first derive more quantitative versions of
Lemmata~\ref{lem-probboundsimple} and \ref{lem-probboundsimple2} by replacing
the input \eqref{eq-uniformupper} and
\eqref{eq-lyaplower} by the estimates \eqref{eq-lyapbound} and
\eqref{eq-lyap}. However, these estimates are not sufficient for the proof of
Theorem~\ref{theo-upper}. In fact, one can
further improve the Lemmata by replacing 
the uniform upper bound  \eqref{eq-lyapbound} by a probabilistic one, deduced
from a large
deviation estimate from \cite{JSS} recalled below and showing
that the transfer matrices grow no more than 
given by the Lyapunov exponent with high
probability. For sake of notational simplicity, we suppose that $E_c=0$.
Furthermore, according to \eqref{eq-lyap} and the Thouless formula 
we may choose positive 
$d<D$ and $\epsilon_0$ such that 
\begin{equation}
\label{eq-lyaplowb}
\gamma(z)
\;\geq\; 
d\,\epsilon^2
\;,
\qquad \mbox{ for }
z\,=\,\epsilon+\imath\delta\; 
\mbox{ with } \;|\epsilon|\,<\,\epsilon_0
\;.
\end{equation}
In order to further simplify notation, we also assume that $\delta,\epsilon>0$
even though all estimates hold with $|\delta|$ and $|\epsilon|$.

\begin{lemma}
\label{lem-probboundimproved}
For $z=\epsilon+\imath\delta$ with $\delta\leq \epsilon<\epsilon_0$ 
introduce the set
\begin{equation}
\label{eq-defupperset}
\Omega_N(z)
\;=\;
\left\{
\,\omega\in\Omega\;\left|\;
\|\Tt_\omega^z(N,0)\|^2\,\geq \,e^{\,d\,\epsilon^2\,N}
\;\right.
\right\}
\;.
\end{equation}
Then
$$
\PP(\Omega_N(z))
\;\geq\;
c_1\;\epsilon
\;.
$$
\end{lemma}

\noindent {\bf Proof.}  This is an immediate corollary of
Lemma~\ref{lem-probboundsimple} with $c_1=d/(2c_0),$ where $c_0$ is
introduced in \eqref{eq-lyapbound}. 
\hfill $\Box$

\begin{lemma}
\label{lem-probboundimproved2}
Let $z=\epsilon+\imath\delta$ with $\delta\leq \epsilon<\epsilon_0$. Then
there exists a constant $c_2$ such that the set
$$   
\hat{\Omega}_N(z)
\;=\;
\left\{
\,\omega\in\Omega\;\left|\;
\max_{0\leq n \leq N}\,\|\Tt_\omega^z(n,0)\|^2
\,\geq\, e^{\epsilon^3\,N}
\;\right.
\right\}
$$
satisfies
$$
\PP(\hat{\Omega}_N(z))
\;\geq\;
1\,-\,e^{-\,c_2\,N}
\;.
$$
\end{lemma}

\noindent {\bf Proof.}
Let us split $N$ into $\frac{N}{N_0}$ pieces of length $N_0$ and follow the
proof of Lemma~\ref{lem-probboundsimple2} invoking 
Lemma~\ref{lem-probboundimproved} instead of Lemma~\ref{lem-probboundsimple},
giving  
$$
\PP\left(
\left\{
\,\omega\in\Omega\;\left|\;
\max_{0\leq n \leq N}\,\|\Tt_\omega^z(n,0)\|^2
\,\geq\, \;e^{\frac{1}{2}\,d\,\epsilon^2\,N_0}
\;\right.
\right\}
\right)
\;\geq\;
1\,-\;(1-c_1\,\epsilon)^{\frac{N}{N_0}}
\;\geq\;
1\,-\;e^{-c_1\,\epsilon\,\frac{N}{N_0}}
\;.
$$
Choosing $N_0=\frac{2\epsilon}d$ shows that one may
take $c_2={c_1d}/{2}$.
\hfill $\Box$

\vspace{.2cm}

Certainly other choices of $N_0$ are possible in the previous proof, but the
present one leading to Lemma~\ref{lem-probboundimproved2}
implies the following estimate, which is sufficient in order to deal 
with one of the
terms in the next section (a boundary term of energies close to $\epsilon_0$).

\begin{coro}
\label{coro-bound1}
Let $z=\epsilon+\imath\delta$ with
$\delta\leq \epsilon<\epsilon_0$. There is a constant $c_3$ such that 
$$
\EE
\left(\,\frac{1}{
\;\max_{0\leq |n| \leq N}\;
\|\,\Tt^z(n,0)\,\|^2\;}
\,\right)
\;\leq\;
e^{-c_3\,\epsilon^3\,N}
\;.
$$
\end{coro}

Now we turn to the refined statements and start by recalling the following 

\begin{theo}
\label{theo-largedeviation} {\rm \cite{JSS}}
Suppose that $\EE( e^{2\imath\eta_\sigma}) \neq 1$ and $0<\alpha\leq
\frac{1}{2}$. 
Then there exist constants $c_4,c_5,c_6,c_7$ such that the set
$$
\Omega_N^\alpha
\;=\;
\left\{
\,\omega\in\Omega\;\left|\;
\max_{0\leq n,m\leq N}\;
\|\Tt_\omega^{\epsilon+\imath\delta}(n,m)\|\,\leq \,e^{c_4}
\;\;\forall\;\;
|\delta|\leq c_5\,N^{-1}\;,\;|\epsilon|\leq N^{-\frac{1}{2}-\alpha}
\right.
\right\}
\;,
$$
satisfies
$$
\PP(\Omega_N^\alpha)
\;\geq\;
1-\,\,c_6\;e^{-c_7\,N^{\alpha}}
\;.
$$
\end{theo}

For a fixed energy $z$, this can be extended to length scales $N$
beyond the localization length (inverse Lyapunov exponent).

\begin{lemma}
\label{lem-improvedupper}
For $z=\epsilon+\imath\delta$ with $\delta\leq c_5\epsilon^2$ and
$\epsilon\geq N^{ -\frac{1}{2}-\alpha}$, the set
$$
\Omega_N^\alpha(z)
\;=\;
\left\{
\,\omega\in\Omega\;\left|\;
\max_{0\leq n,m\leq N}\;
\|\Tt_\omega^{z}(n,m)\|\,\leq \,e^{c_4\,N\,
\epsilon^{2(1-2\alpha)}}
\right.
\right\}
\;,
$$
satisfies
$$
\PP(\Omega_N^\alpha(z))
\;\geq\;
1-\,\,c_6\,N\,\epsilon\;e^{-c_7\,\epsilon^{-\alpha}}
\;.
$$
\end{lemma}

\noindent {\bf Proof.} 
We split $N$ into $\frac{N}{N_0}$ pieces of length
$N_0=\epsilon^{-\frac{2}{2\alpha+1}}$. The condition 
$\epsilon\geq N^{ -\frac{1}{2}-\alpha}$ insures that $N_0\leq N$. By
the stationarity, on each
piece we may apply Theorem~\ref{theo-largedeviation} (with $N_0$ instead of
$N$) because $\epsilon\leq
N_0^{-\frac{1}{2}-\alpha}$. For the $j$th piece, denote the set appearing in 
Theorem~\ref{theo-largedeviation}  by $\Omega_N^{\alpha,j}$. For any
$\omega\in \cap_{j=1,\ldots,\frac{N}{N_0}} \Omega_N^{\alpha,j}$ one then has
for $\delta\leq c_5N_0^{-1}$ (and hence also $\delta\leq c_5\epsilon^2$) 
the estimate
$$
\max_{0\leq n,m\leq N}\;
\|\Tt_\omega^{z}(n,m)\|
\;\leq \;
e^{c_4\,\frac{N}{N_0}}
\;\leq \;
e^{c_4\,N\,
\epsilon^{2(1-2\alpha)}}
\;.
$$
Therefore $\cap_{j=1,\ldots,\frac{N}{N_0}} \Omega_N^{\alpha,j}\subset
\Omega_N^\alpha(z)$. We hence deduce
from Theorem~\ref{theo-largedeviation} that
$$
\PP(\Omega_N^\alpha(z)^c)
\;\leq\;
c_6\,\frac{N}{N_0}\;e^{-c_7\,N_0^{\alpha}}
\;\leq\;
c_6\,N\,\epsilon \;e^{-c_7\,\epsilon^{-\alpha}}
\;,
$$
which is precisely the statement of the lemma.
\hfill $\Box$

\vspace{.2cm}

This last lemma can now be used in order to improve
Lemma~\ref{lem-probboundimproved} in the range $\delta <c\epsilon^2.$ 

\begin{lemma}
\label{lem-probboundoptimal}
Let $z=\epsilon+\imath\delta$ with
$\delta\leq c_5\epsilon^2$ and
$N^{ -\frac{1}{2}-\alpha}\leq
\epsilon\leq N^{-\alpha}, \; 0<\alpha\leq \frac 12$. Then the set
$\Omega_N(z)$ defined in {\rm \eqref{eq-defupperset}} satisfies
for some constant $c_8$
$$
\PP(\Omega_N(z))
\;\geq\;
c_8\;\epsilon^{4\alpha}
\;.
$$
\end{lemma}

\noindent {\bf Proof.} 
We argue as in the proof of Lemma~\ref{lem-probboundsimple}.
Let us set again $P=\PP(\Omega_N(z))$, and estimate separately the
contribution from the
complement of $\Omega_N(z),$  $\Omega^\alpha_N(z)$ and it's complement. Due to \eqref{eq-lyaplowb}, 
Lemma~\ref{lem-improvedupper} and the a priori bound 
\eqref{eq-lyapbound} (used on the complement of $\Omega^\alpha_N(z)$), 
it follows that
$$
d\,\epsilon^2
\;\leq\;
(1-P)\,\frac{1}{2}\,d\,\epsilon^2
\;+\;
P\,c_4\,
\epsilon^{2(1-2\alpha)}
\;+\;
2\,c_0\,c_6\,N\,\epsilon^2\;e^{-c_7\,\epsilon^{-\alpha}}
\;.
$$
Hence
$$
P
\;\geq\;
\epsilon^{4\alpha}\;
\frac{d-4\,c_0\,c_6\,N\,e^{-c_7\,\epsilon^{-\alpha}}}{
2\,c_4-d\,\epsilon^{4\alpha}}
\;.
$$
The hypothesis $\epsilon\leq N^{-\alpha}$ implies the result
(it would be enough to assume $\epsilon\leq \log(N)^p$ for some $p$).
\hfill $\Box$

\vspace{.2cm}

As the final preparatory step for the next section, we improve 
Lemma~\ref{lem-probboundimproved2} in the range $\delta <c\epsilon^2,$
by invoking
Lemma~\ref{lem-probboundoptimal} in its proof.

\begin{lemma}
\label{lem-probboundoptimal2}
Let $z=\epsilon+\imath\delta$ with $\delta\leq c_5\epsilon^2$ and
$ N^{ -\frac{1}{2}-\alpha}\leq\epsilon\leq N^{-\alpha}, \; 0<\alpha\leq \frac 12$.
Then the set
$$
\hat{\Omega}^\alpha_N(z)
\;=\;
\left\{
\,\omega\in\Omega\;\left|\;
\max_{0\leq n \leq N}\,\|\Tt_\omega^z(n,0)\|^2
\,\geq\,  e^{N^{1-\alpha}\epsilon^{2(1+2\alpha)}}
\;\right.
\right\}
$$
satisfies for some constant $c_9$
$$
\PP(\hat{\Omega}^\alpha_N(z))
\;\geq\;
1\,-\,e^{-\,c_9\,N^{\alpha}}
\;.
$$
\end{lemma}

\noindent {\bf Proof.}
Splitting $N$ into $\frac{N}{N_0}$ pieces of length $N_0$ and arguing
exactly as in Lemma~\ref{lem-probboundsimple2} invoking 
Lemma~\ref{lem-probboundoptimal} instead of
Lemma~\ref{lem-probboundsimple}, we obtain 
$$
\PP\left(
\left\{
\,\omega\in\Omega\;\left|\;
\max_{0\leq n \leq N}\,\|\Tt_\omega^z(n,0)\|^2
\,\geq\, \;e^{\frac{1}{2}\,d\,\epsilon^2\,N_0}
\;\right.
\right\}
\right)
\;\geq\;
1\,-\;(1-c_8\,\epsilon^{4\alpha})^{\frac{N}{N_0}}
\;.
$$
Choosing $N_0=(2N^{1-\alpha}\epsilon^{4\alpha})/d$ shows that one may
take $c_9={c_8d}/{2}$.
\hfill $\Box$

\vspace{.2cm}

Lemma~\ref{lem-probboundoptimal2} 
implies the following estimate, which is the main result of this section and
will be used in the next one.

\begin{coro}
\label{coro-bound2}
Let $z=\epsilon+\imath\delta$. 
If $\delta\leq c_5\epsilon^2$ and
$ N^{ -\frac{1}{2}-\alpha}\leq\epsilon\leq N^{-\alpha}, \; 0<\alpha\leq \frac 12$, one has
$$
\EE
\left(\,\frac{1}{
\;\max_{0\leq |n| \leq N}\;
\|\,\Tt^z(n,0)\,\|^2\;}
\,\right)
\;\leq\;
e^{-N^{1-\alpha}\epsilon^{2(1+2\alpha)}}
\;+\;
e^{-\,c_9\,N^{\alpha}}
\;.
$$
\end{coro}

\section{Proof of upper bound for random polymer models} 
\label{5}

In this section we complete the proof of Theorem~\ref{theo-upper}. For this
purpose, we  follow the strategy discussed in Section~\ref{sec-uppergen} and
consider $M^{q,\alpha_0,\alpha_1}_T(E_0,E_1)$ 
defined as in \eqref{eq-greenpart}, but
with a disorder average $\EE$.
By Proposition~\ref{prop-CT} it is sufficient to bound
$M^{q,0,1+\alpha}_T(E_0,E_1)$ for $\alpha>0$ if $(E_0,E_1)$ contains the
spectrum. Moreover, energies bounded
away from critical energies have a strictly positive Lyapunov exponent
\cite{BG}. By the results of Section~\ref{sec-logbound}, 
these energies hence  lead at most to logarithmic growth in time, and
therefore give no contribution to the diffusion exponents $\beta_q$. Thus we
are left to deal with energy intervals around the critical energies. All of
them are treated the same way, so we focus on one of them. We suppose that
$E_c=0$ and consider only the energy interval $[0,\epsilon_0]$ with
$\epsilon_0$ chosen as in \eqref{eq-lyaplowb}; the other side
$[-\epsilon_0,0]$ is again treated similarly. Furthermore we split the
contribution as follows 
$$
M^{q,0,1+\alpha}_T(0,\epsilon_0)
\;=\;
M^{q,0,1+\alpha}_T(0,T^{-\eta})
\;+\;
M^{q,0,1+\alpha}_T(T^{-\eta},T^{-\alpha})
\;+\;
M^{q,0,1+\alpha}_T(T^{-\alpha},\epsilon_0)
\;,
$$
where $\eta=\min\{q,\frac{1}{2}\}$. This is a good choice due 
to the following lemma, showing that the contribution
$M^{q,0,1+\alpha}_T(0,T^{-\eta})$ is bounded by the diffusion exponent as
given in Theorem~\ref{theo-upper}.

\begin{lemma}
\label{lem-centralpart}
For some constant $C_1$,  one has
$$
M^{q,0,1+\alpha}_T(0,T^{-\eta})
\;\leq\;
C_1\,T^{q-\eta+\alpha q}
\;.
$$
\end{lemma}

\noindent {\bf Proof.}
By (\ref{m1a}) we have 
$$
M^{q,0,1+\alpha}_T(0,T^{-\eta})
\;\leq\;
T^{q(1+\alpha)}\,
\int^{T^{-\eta}}_0d\epsilon\;
\Im m\,B_\Nn(\epsilon+\imath \,T^{-1})
\;.
$$
The estimate now follows from \eqref{eq-DOS} and
Proposition~\ref{prop-Borel} in the appendix, with $\epsilon_0\,=\,T^{-\eta}.$
\hfill $\Box$

\vspace{.2cm}

Next let us consider the boundary term
$M^{q,0,1+\alpha}_T(T^{-\alpha},\epsilon_0)$. 

\begin{lemma}
\label{lem-boundarypart}
For some constant $C_2=C_2(\alpha)$,  one has
$$
M^{q,0,1+\alpha}_T
(T^{-\alpha},\epsilon_0)
\;\leq\;
T^{4\,\alpha\,q}\;+\;C_2
\;.
$$
\end{lemma}

\noindent {\bf Proof.} Let us split the sum over $n$ appearing in the
definition of $M^{q,0,1+\alpha}_T
(T^{-\alpha},\epsilon_0)$ into one over $|n|\leq T^{4\alpha}$ and the other
over $|n|> T^{4\alpha}$. The first one can be bounded by $T^{4\alpha q}$
using \eqref{m1b}. In the second one we apply
Proposition~\ref{prop-DT} combined with Corollary~\ref{coro-bound1} in order
to bound the Green's function. This shows $M^{q,0,1+\alpha}_T
(T^{-\alpha},\epsilon_0)$ is bounded above by
$$
T^{4\,\alpha\, q}
\;+\;
\int^{\epsilon_0}_{T^{-\alpha}}\;
\frac{d\epsilon}{\pi\,T}\;
\sum_{|n|>\,T^{4\,\alpha}}
\;|n|^q\;16\;\tau^4\;T^6\,
e^{-c_3\,\epsilon^3\,|n|}
\;.
$$ 
 In the second term, the
sum over $n$ is bounded by Lemma~\ref{lem-sumbound}. As
$\epsilon>T^{-\alpha}$, it follows that the second term is bounded by
$CT^pe^{-c_3T^{\alpha}}$ for some $C,p>0$. Hence this term gives the second
contribution in the bound. 
\hfill $\Box$

\vspace{.2cm}

\begin{lemma}
\label{lem-mainpart}
For $\eta=\min\{q,\frac{1}{2}\}$ 
and some constants $C_3=C_3(\alpha)$ and $C_4$,  one has
$$
M^{q,0,\infty}_T
(T^{-\eta},T^{-\alpha})
\;\leq\;
C_3\;+\;C_4\,T^{6q\alpha}\,\cdot\,
\left\{
\begin{array}{cc}
T^{q-\frac{1}{2}} & \;\;\;\; q\geq \frac{1}{2}\;,
\\
&
\\
T^{\alpha(2q-1)} & \;\;\;\; q\leq \frac{1}{2}\;.
\end{array}
\right.
$$
\end{lemma}

\noindent {\bf Proof.}
We first split
$M^{q,0,\infty}_T(T^{-\eta},T^{-\alpha})$ 
into two contributions $M_T(1)$ and $M_T(2)$, the first containing all the
summands with $|n|$ smaller than the (energy dependent) localization length:
$$
M_T(1)
\;=\;
\int^{T^\alpha}_{T^{-\eta}} \frac{d\epsilon}{\pi\,T}\;
\sum_{1\leq |n|\leq \epsilon^{-2} T^{4\alpha}}
\,|n|^q
\;\EE\;|G_\omega^{\epsilon+\frac{\imath}{T}}(0,n)|^2
\;,
$$
and the second $M_T(2)$ containing the sum over $|n|>\epsilon^{-2} T^{4\alpha}$
corresponding to the summands beyond the localization length. $M_T(1)$
is bounded using \eqref{m1a}:
$$
M_T(1)
\;\leq\;
T^{4\,\alpha\, q}\,
\int \Nn(dE)\;
\int_{T^{-\eta}}^{T^{-\alpha}}\;
\frac{d\epsilon}{\pi\,T}\;
\epsilon^{-2\,q}
\;\frac{1}{(E-\epsilon)^2+T^{-2}}
$$
In order to bound the factor $\epsilon^{-2q}$, let us split the integral over
$\epsilon$ into $\frac{\eta-\alpha}{\alpha}$ pieces:
\begin{eqnarray*}
M_T(1)
& \leq &
T^{4\,\alpha\, q}\,
\sum_{j=1}^{\frac{\eta-\alpha}{\alpha}}
\;T^{2q(\eta-(j-1)\alpha)}\;
\int \Nn(dE)\;
\int_{T^{-\eta+(j-1)\alpha}}^{T^{-\eta+j\alpha}}\;
\frac{d\epsilon}{\pi\,T}
\;\frac{1}{(E-\epsilon)^2+T^{-2}}\\ 
& = & 
T^{4\,\alpha\, q}\,
\sum_{j=1}^{\frac{\eta-\alpha}{\alpha}}
\;T^{2q(\eta-(j-1)\alpha)}\;\int_{T^{-\eta+(j-1)\alpha}}^{T^{-\eta+j\alpha}}
d\epsilon\;
\Im m\;
B_\Nn(\epsilon+\imath\,T^{-1})
\;.
\end{eqnarray*}
Using Proposition~\ref{prop-Borel} we obtain
$$
M_T(1)
\;\leq\;
C\,T^{4\,\alpha\, q}\,
\sum_{j=1}^{\frac{\eta-\alpha}{\alpha}}
\;T^{2q(\eta-(j-1)\alpha)}
\;T^{-\eta+j\alpha}\;
\;=\;
C\,
T^{6q\alpha}\,T^{(2q-1)\eta}
\,\sum_{j=1}^{\frac{\eta-\alpha}{\alpha}}
\;T^{(1-2q)\alpha\,j}  
\;.
$$
For $q\geq 1/2$, we use the
bound $T^{(1-2q)\alpha\,j}\leq 1$ showing that the sum is bounded by  
$\frac{\eta-\alpha}{\alpha}$. For $q\leq 1/2$ we bound each summand by
$T^{(1-2q)\alpha\,j}\leq T^{(1-2q)(\eta-\alpha)}$. This gives the second
contribution in the lemma.

\vspace{.1cm}

It remains to show that $M_T(2)\leq C_3$. Due to
Proposition~\ref{prop-DT} and Corollary~\ref{coro-bound2},
$$
M_T(2)
\;\leq\;
\int_{T^{-\eta}}^{T^{-\alpha}}\;
\frac{d\epsilon}{\pi\,T}\;
\sum_{|n|>\,\epsilon^{-2}\,T^{4\,\alpha}}
\;|n|^q\;16\;\tau^4\;T^6\,
\left(
e^{-\epsilon^{2(1+2\alpha)}\,|n|^{1-\alpha}}
\,+\,
e^{-c_9\,|n|^\alpha}
\right)
\;.
$$
Using Lemma~\ref{lem-sumbound} it is now elementary to
bound $M_T(2)$ by a constant.
\hfill $\Box$

\vspace{.2cm}

Combining Lemmata \ref{lem-centralpart}, 
\ref{lem-boundarypart} and \ref{lem-mainpart}, and recalling that
$\alpha$ can be taken arbitrary close to $0,$  proves
Theorem~\ref{theo-upper}.

\section*{Appendix: estimates on the Borel transform} 
\label{appendix}

In Section~\ref{5} we used well-known 
estimates on the Borel transform $B_\mu(z)=\int\mu(dE) \,(z-E)^{-1}$ 
of a measure $\mu$.
For sake of completeness we provide a short proof.

\begin{proposi}
\label{prop-Borel}
If a measure $\mu$ satisfies at some $E$ the bound
$\mu([E-\epsilon,E+\epsilon])<C\epsilon$ for all $\epsilon>0$,
then 
for any  finite positive $\delta$ and $\epsilon_0$
$$
\Im m\, B_\mu(E+\imath\delta)\;<\;\frac{\pi}{2}\;C\;,
\qquad
\int_0^{\epsilon_0}d\epsilon\;\Im m \,
B_\mu(E+\epsilon+\imath\delta)
\;<\;\pi^2\,C\,\epsilon_0
\;.
$$
\end{proposi}

\noindent {\bf Proof.}
One has, uniformly in $\delta$,
\begin{eqnarray*}
\Im m\,
B_\mu(E+\imath\delta)
& = & 
\int
\mu(de)\;
\frac{\delta}{(e-E)^2+\delta^2}
\;=\;
\delta
\int_{0}^{\frac 1{\delta^2}}dt\; 
\mu\left(\left\{e\in\RR\,\left|
\;|e-E|<\sqrt{\frac 1t -\delta^2}
\right.\right\}\right)
\\
& < &
C\,\delta\,
\int_{0}^{\frac 1{\delta^2}}dt\;\sqrt{\frac 1t
  -\delta^2}
\;<\;
C\int_0^\infty dx\;\frac{\sqrt{x}}{(x+1)^2}
\;,
\end{eqnarray*}
which gives the first inequality. For the second one, let us bound
the indicator function  $\chi_{[-\epsilon_0,\epsilon_0]}(\epsilon)$  above by
$\frac {2\epsilon_0^2}{\epsilon^2+\epsilon_0^2}$. Then one can use 
the stability of Cauchy distribution to obtain
\begin{eqnarray*}
\int_0^{\epsilon_0}d\epsilon\;
\Im m\, B_\mu(E+\epsilon+\imath\delta)
& < & 
\int d\epsilon\;\frac {2\,\epsilon_0^2}{\epsilon^2+\epsilon_0^2}\;
\int\mu(de)\;
\frac{\delta}{(E+\epsilon-e)^2+\delta^2}
\\
& = &  2\,\pi\,\epsilon_0\,\int\mu(de)\;
\frac{\delta+\epsilon_0}{(E-e)^2+(\delta+\epsilon_0)^2}\;,
\\
& = &\;2\,\pi\epsilon_0\,\Im m\,B_\mu(E+\imath(\delta+\epsilon_0))\;,
\end{eqnarray*}
and thus the second inequality follows from the first one.
\hfill $\Box$


\end{document}